\documentclass[pre,aps,twocolumn]{revtex4}
\usepackage{amsfonts,amssymb,amsmath,bm}
\usepackage{graphicx,color}

\begin{document}

\title{Two-dimensional turbulence in a finite box}

\author{I.V. Kolokolov$^{1,2}$, V.V. Lebedev$^{1,2}$}

\affiliation{$^1$Landau Institute for Theoretical Physics, RAS, \\
142432, Chernogolovka, Moscow region, Russia; \\
$^2$National Research  University Higher School of Economics, \\
101000, Myasnitskaya ul. 20, Moscow, Russia.}

\date{\today}

\begin{abstract}

We present theory of two-dimensional turbulence excited by an external force in thin fluid films on scales larger than the film thickness. The principal feature of two-dimensional turbulence is the tendency of producing motions of larger and larger scales thanks to the nonlinear interaction. The tendency leads to formation of the so-called inverse cascade and, at some conditions, of big coherent vortices. We discuss the mean velocity profile of the coherent vortices and the flow fluctuations on the background of the mean velocity for different regimes. We demonstrate that the regime of strongly interacting fluctuations leads to an anisotropic scaling inside the coherent vortices.

\end{abstract}

\maketitle

\section{Introduction}
\label{sec:intro}

It is a great honor for us to present this work for a collection of articles dedicated to the 130th anniversary of P.L. Kapitsa. Pyotr Leonidovich's discovery of superfluidity significantly expanded our notion of possible hydrodynamic effects. Hydrodynamics still continues to amaze us with the variety of phenomena related to it. In this article we present the current state of the theory of two-dimensional turbulence, which has been actively developing in recent years.

We consider turbulent states of thin fluid films characterized by an intense chaotic motions of the fluid. On scales larger than the film thickness, the hydrodynamic motions in such films can be treated as two-dimensional \cite{BE12}. Other systems where effectively two-dimensional hydrodynamic motions can be produced are suspended soap or smectic films \cite{kellay2002two,Yablonskii2017}. Note also the effectively two-dimensional subsystem of the fast rotating three-dimensional fluid \cite{Proudman,Pedlovsky}. The turbulent states in the above systems differ essentially from the three-dimensional turbulence.

Turbulence can be excited by a permanently acting force or evolve without external forcing. The first case is called forced turbulence and the second case is called decaying turbulence. The decaying case is in many aspects analogous to the pumping case. However, the decaying case has some peculiarities, that lie without scope of our paper. We examine the forced case. In our setup, the external force can be static or can be chaotic function of time. In the second case we assume, that the force has stationary statistical properties. In addition, one can consider a force periodically varying in time. In all the cases the turbulent state, being a state with strong fluctuations, has stationary statistical properties.

Already first theoretical works \cite{67Kra,68Lei,69Bat} devoted to two-dimensional turbulence reveal its principal difference from three-dimensional one. From the theoretical point of view, the difference is related to existence of two quadratic positively defined quantities (energy and enstrophy) conserved by two-dimensional Euler equation. That leads to forming two different cascades produced by the non-linear interaction: enstrophy is carried from the pumping scale to smaller scales (direct cascade) whereas energy is carried to larger scales (inverse cascade). The enstrophy is dissipated due to viscosity at scales smaller than the pumping length and the energy is dissipated due to bottom friction at scales larger than the pumping length.

Statistical properties of the velocity fluctuations in the inverse cascade were investigated both experimentally \cite{Tabeling02} and numerically \cite{BCV}. Results of the works are in a good agreement with the analytical theory developed for an unbounded system \cite{KraM}. Amazingly, the normal Kolmogorov scaling is observed in the inverse cascade \cite{BCV}. The normal scaling in $2d$ is in contrast with the anomalous scaling observed in three-dimensional turbulence \cite{Frisch}. The direct (enstrophy) cascade is characterized by its own scaling laws \cite{71Kra,94FL}.

For a three-dimensional flow in a fluid, the only dissipation mechanism of the kinetic energy is viscosity. Thus, there is the only dimensionless parameter, Reynolds number, characterizing the degree of non-linearity for an incompressible flow. In contrast, at considering effectively two-dimensional thin fluid films one deals with two dissipative mechanisms: viscosity and bottom friction. {Of course, bottom friction is reduced to the same viscosity, however, acting at the scales of the order of the film thickness. Therefore at the two-dimensional analysis bottom friction has to be taken into account separately.} An interplay of the dissipative mechanisms leads to a more complicated characterization of the flow nonlinearity in two-dimensional turbulence.

The traditional way to produce flow in $2d$ hydrodynamics is applying to the fluid an external periodic in space static force (Kolmogorov force). Such setup is used both in laboratory experiments with thin liquid films \cite{Sommeria86,Tabeling02,Shats09,Orlov17} and in numerical simulations \cite{MCH04,Mishra15,DFKL22}. Then the transition to turbulence appears to be complicated, it could be soft transition or jump (depending on the ratio between the pumping length and the box size) and could go through some intermediate stages. Another possibility, realized in many numerical simulations, is the random force short correlated in time \cite{BCV,Bof07,Laurie2014,frishman17,Parfenyev2022,Xu}.

For an intensive pumping, the inverse cascade, carrying the energy upscales, leads to an accumulation of the energy at the box size. The accumulation can lead to formation of coherent vortices of size comparable with the box size. The coherent vortices are long-living and possess well determined mean velocity profile. As numerics and experiment show, the profile is isotropic and corresponds to a differential rotation. As it was first demonstrated in the work \cite{Laurie2014}, where the model of short correlated in time pumping force was numerically examined, the velocity profile is flat, that is the (polar) velocity of the profile is independent of the distance to the center of the vortex. Further the observation was confirmed numerically \cite{Parfenyev2022} and experimentally \cite{Orlov17} and some theoretical argument in favor of the flat profile were formulated \cite{KL15,kolokolov2016structure,kolokolov2016velocity,KL17}. Conditions needed for appearing the coherent vortices are discussed in Ref. \cite{DFKL22}.

The plan of our subsequent text is as follows. In Section \ref{sec:general} we present basic relations concerning two-dimensional hydrodynamic flows. In Section \ref{sec:unbounded} we observe properties of two-dimensional turbulence in an unbounded fluid film. In Section \ref{sec:coherent} we introduce equations describing the coherent vortex. In Section \ref{sec:quasilin} we analyze fluctuations inside the coherent vortex in the so-called quasilinear regime. In Section \ref{sec:strong} we examine the regime of strongly interacting fluctuations inside the coherent vortex. In Section \ref{sec:conclu} we outline our results and discuss possible perspectives and relations to other physical phenomena.

\section{General relations}
\label{sec:general}

A two-dimensional fluid flow is described by its (two-dimensional) velocity field $\bm v$, it is a function of time and two coordinates. We assume that the flow is incompressible: $\nabla \bm v=0$. The property implies that the Mach number $v/c$ is small, where $c$ is the speed of sound. This condition is usually satisfied well in really excited turbulent flows. In the section we formulate basic equations describing two-dimensional incompressible hydrodynamics.

Having in mind dynamics of thin films, one uses two-dimensional Navier-Stokes equation, supplemented by the term related to bottom friction: 
\begin{equation}
\partial_t \bm v +(\bm v \nabla) \bm v +\nabla p
=-\alpha \bm v +\nu \nabla^2 \bm v +\bm f .
\label{basic0}
\end{equation}
Here $\bm f$ is the external force (per unit mass), $p$ is pressure (per unit mass), $\nu$ is the kinematic viscosity coefficient and $\alpha$ is the bottom friction coefficient. Taking divergence of the equation (\ref{basic0}) and using the incompressibility condition $\nabla \bm v=0$, we arrive at the following relation for the pressure field $p$:
\begin{equation}
\nabla^2 p=-(\partial_\lambda v_\mu)(\partial_\mu v_\lambda),
\label{pressure}
\end{equation}
where Greek subscripts, running over $1,2$, designate vector components along two orthogonal axes. At deriving Eq. (\ref{pressure}) we implied the condition $\nabla \bm f=0$.

We assume that the pumping force $\bm f$ has the characteristic wave vector $k_f$. Then $\nu k_f^2$ is the viscous damping at the pumping scale. It is instructive to compare the damping with $\alpha$. For thin liquid films it is easy to achieve the inequality $\alpha\gg \nu k_f^2$, since $\alpha$ can be estimated as $\nu h^{-2}$, where $h$ is the thickness of the film. The characteristic length of the pumping force should be larger than $h$ (otherwise the flow cannot be treated as two-dimensional one), thus $k_f h\gtrsim 1$. That leads to the conclusion $\alpha \gtrsim \nu k_f^2$.

However, to observe effects related, say, to intensive large-scale vortices in the turbulent regime, one should make $\alpha$ as small as possible. To achieve the goal some experimental tricks are used related to multilayered films \cite{Tabeling02,Shats09,Orlov17}. Note also, that $\alpha$ is small compared to $\nu k_f^2$ for the two-dimensional subsystem of the fast rotating three-dimensional fluid \cite{Proudman,Pedlovsky}, where the effective bottom friction coefficient is related to the convective motion accompaning by formation of Ekman boundary layers \cite{Vergeles21}. Other cases, where small $\alpha$ can be realized, include suspended soap or smectic films, which have no contact with a solid boundary \cite{kellay2002two,Yablonskii2017}.

In two dimensions, it is convenient to describe the flow in terms of vorticity $\varpi$, defined as
\begin{equation}
\varpi= \mathrm{curl}\, \bm v
\equiv \partial_1 v_2-\partial_2 v_1.
\label{varpi}
\end{equation}
Obviously, vorticity $\varpi$ is a scalar (or, more precisely, a pseudoscalar) field. The equation controlling the vorticity field is derived from Eq. (\ref{basic0}) by taking curl of it:
\begin{equation}
\partial_t\varpi+ \bm v \nabla \varpi
= -\alpha \varpi +\nu \nabla^2 \varpi +\phi,
\label{basic}
\end{equation}
where $\phi =\mathrm{curl}\, \bm f$. Note that pressure $p$ drops from the equation (\ref{basic}) for vorticity.

To close the equation (\ref{basic}) one should restore the velocity field $\bm v$ from the vorticity field $\varpi$. Due to the incompressibility condition $\partial_1 v_1+\partial_2 v_2=0$, it is possible to introduce the stream function $\psi$, related to the velocity components and to the vorticity as
\begin{equation}
v_1={\partial_2}\psi, \quad
v_2=-{\partial_1}\psi, \quad
\varpi=-\nabla^2 \psi.
\label{stre1}
\end{equation}
To express the stream function in terms of $\varpi$, it is necessary to solve the Laplace equation $\nabla^2 \psi=-\varpi$. Generally, the equation should be solved with suitable boundary conditions. For an unbounded system, one can exploit the relation
\begin{equation}
\psi(\bm r)=-\frac{1}{2\pi}
\int d^2 x\, \varpi(\bm x) \ln |\bm r-\bm x|.
\label{stre2}
\end{equation}
After calculating the integral one finds the velocity components in accordance with Eq. (\ref{stre1}), thus expressing the velocity via the vorticity.

For weak external forcing the nonlinear term $(\bm v \nabla)\bm v$ in the equation (\ref{basic0}) can be neglected and we arrive at the linear equation. In terms of vorticity, the equation turns to
\begin{equation}
(\partial_t+\alpha -\nu \nabla^2)\varpi=\phi.
\label{basiclin}
\end{equation}
It describes a laminar flow, excited by the external forcing. Due to the presence of dissipation (viscosity and bottom friction) after some transition processes the flow becomes stationary, if the force $\bm f$ is stationary. If the force $\bm f$ is random with statistics stationary in time then the velocity $\bm v$ in the laminar regime will be random with statistics stationary in time as well.

An influence of the external force to the flow can be characterized by its average power (per unit mass) $\epsilon = \langle \bm f \bm v \rangle$, where angular brackets mean averaging over time. The quantity $\epsilon$ is often called the energy flux. Generally, the average $\langle \bm f \bm v \rangle$ depends on coordinates. We assume that the dependence is smooth, that is its characteristic length is larger than characteristic scales of turbulent pulsations. Then the inhomogeneity of $\epsilon$ is irrelevant for the theory.

A simple analysis of the role of the nonlinearity, based on dimension estimates, shows that there are two dimensionless parameters characterizing the strength of the hydrodynamic non-linear interaction
\begin{equation}
\beta_\nu=\frac{\epsilon}{\nu^{3} k_f^{4}}, \qquad
\beta_\alpha=\frac{\epsilon k_f^2}{\alpha^3},
\label{smallp}
\end{equation}
where $k_f$ is the characteristic wave vector of the pumping force $\bm f$. The first parameter $\beta_\nu$ in Eq. (\ref{smallp}) is a power of Reynolds number taken at the pumping scale $k_f^{-1}$. The second parameter $\beta_\alpha$ in Eq. (\ref{smallp}) is related to bottom friction and characterizes its role at the same pumping scale.

If the inequalities $\beta_\alpha\gg1$ and $\beta_\nu\gg1$ are satisfied then developed turbulence is excited and pulsations of different scales are formed due to non-linear interaction of the flow fluctuations. The energy produced by the forcing at the scale $k_f^{-1}$ flows to larger scales whereas the enstrophy produced by the forcing at the scale $k_f^{-1}$ flows to smaller scales \cite{67Kra,68Lei,69Bat}. Thus two cascades are formed: the energy cascade (inverse cascade), realized at scales larger than the forcing scale $k_f^{-1}$, and the enstrophy cascade (direct cascade) realized at scales smaller than the forcing scale $k_f^{-1}$.

\section{Turbulence in an unbounded system}
\label{sec:unbounded}

Here we consider two-dimensional turbulence in a big box, where its finite size do not play an essential role. To describe turbulence in the case one can use the model of an isotropic and homogeneous in space and time turbulence. Of course, then the pumping force should be isotropic and homogeneous in space and time as well, at least in the statistical sense.

Multiplying the dynamic equation (\ref{basic0}) by $\bm v$ and averaging it, one obtains the energy balance
\begin{equation}
\epsilon\equiv \langle \bm f \bm v \rangle
= \alpha \langle v^2 \rangle
+\nu \langle (\nabla_\alpha \bm v)^2 \rangle.
\label{energyb}
\end{equation}
Remind that the angular brackets mean averaging over time or, in the theoretical framework, averaging over the system statistics. At deriving Eq. (\ref{energyb}) we omitted all full derivatives over time and coordinates, having in mind homogeneity in space and time. The relation (\ref{energyb}) has simple physical meaning: the energy pumped into the fluid by the external force is dissipated by bottom friction and viscosity.

Analogously, multiplying the equation (\ref{basic}) by $\varpi$ and averaging, one obtains one extra balance equation
\begin{equation}
\eta\equiv\langle \phi \varpi \rangle=
\alpha \langle \varpi^2 \rangle
+\nu \langle (\nabla\varpi)^2\rangle,
\label{enstrophyb}
\end{equation}
Again, at deriving Eq. (\ref{enstrophyb}) we omitted all full derivatives over time and coordinates, having in mind homogeneity in space and time. The quantity $\eta$ is often called the enstrophy flux, it can be estimated as $\eta\sim \epsilon k_f^2$. The relation (\ref{enstrophyb}) has simple physical meaning: the enstrophy pumped into the fluid is dissipated by bottom friction and viscosity.

One can derive important relations for the energy and the enstrophy cascades following the Kolmogorov scheme \cite{Kol41,LL87}. In the case of isotropic homogeneous in space turbulence one finds for the inverse energy cascade \cite{67Kra,68Lei,69Bat}
\begin{equation}
\left\langle \left\{\frac{\bm r}{r}
\left[\bm v(\bm r_1)-\bm v(\bm r_2)\right]
\right\}^3 \right\rangle=\frac{3}{2}\epsilon r,
\label{inverseflux}
\end{equation}
where $\bm r=\bm r_1-\bm r_2$ and angular brackets, as above, mean averaging over time. At the same conditions one finds for the direct enstrophy cascade
\begin{equation}
\left\langle \frac{\bm r}{r}
\left[\bm v(\bm r_1)-\bm v(\bm r_2)\right]
\left[ \varpi(\bm r_1)-\varpi(\bm r_2)\right]^2
\right\rangle=-2\eta r.
\label{directflux}
\end{equation}
Here $r\ll k_f^{-1}$. Note the opposite signs in the right hand sides of Eqs. (\ref{inverseflux},\ref{directflux}) reflecting opposite directions of the energy and of the enstrophy flows.

Based on the relation (\ref{inverseflux}) one can formulate the following estimate for the velocity differences in the inverse cascade
\begin{equation}
|\bm v(\bm r_1)-\bm v(\bm r_2)|
\sim (\epsilon r)^{1/3},
\label{inversec}
\end{equation}
where $r=|\bm r_1-\bm r_2|$. Validity of the simple estimate (\ref{inversec}) is confirmed by laboratory experiments and numerical simulations \cite{Tabeling02,BCV}. It is a difference between $2d$ turbulence and $3d$ turbulence, where the anomalous scaling is observed \cite{Frisch}. For the direct cascade the estimate for the velocity differences, based on Eq. \ref{directflux}), reads
\begin{equation}
|\bm v(\bm r_1)-\bm v(\bm r_2)|
\sim (\epsilon k_f^2)^{1/3} r.
\label{directc}
\end{equation}
The estimate (\ref{directc}) is correct upto logarithmic corrections \cite{71Kra,94FL}.

Comparing the nonlinear term and the term with the bottom friction in Eq. (\ref{basic0}), we find from Eq. (\ref{inversec}) that the energy cascade is terminated by the bottom friction at the scale $\sim L_\alpha$,
 \begin{equation}
 L_\alpha = \epsilon^{1/2}\alpha^{-3/2}.
 \label{Lalpha}
 \end{equation}
Analogously, comparing the nonlinear term and the viscous one in Eq. (\ref{basic0}) and using the estimate (\ref{directc}) we conclude that the enstrophy cascade is terminated by viscosity at the scale $\sim L_\nu$,
\begin{equation}
L_\nu = \nu^{1/2} \eta^{-1/6} \sim  \nu^{1/2} (\epsilon k_f^2)^{-1/6}.
\label{Lnu}
\end{equation}
The inequalities $\beta_\alpha\gg1$, $\beta_\nu\gg1$ are equivalent to the inequalities $k_f L_\alpha \gg1$ and $k_f L_\nu \ll1$. Thus there are regions above and below the pumping scale $k_f^{-1}$, where the inverse and the direct cascades are realized.

In accordance with the estimates (\ref{inverseflux},\ref{directflux}) the velocity gradient is independent of the scale in the direct cascade and diminishes as the scale increases in the inverse cascade. Besides, the velocity fluctuations increase as the scale increases in the inverse cascade. That is why in the energy balance (\ref{energyb}) the second term in the right hand side is negligible, it is explained by the condition $k_f L_\alpha \gg1$. By other words, energy is dissipated mainly by bottom friction at the scale $L_\alpha$. Inversely, enstrophy is dissipated mainly by viscosity at the small scale $L_\nu$. Therefore the first term in the right hand side of Eq. (\ref{enstrophyb}) is negligible, it is explained by the condition $k_f L_\nu \ll1$.

Let us formulate the criteria of the applicability of the relations given in this section. We established that the turbulent pulsations have scales between $L_\alpha$ and $L_\nu$. Thus, for validity of the model of the unbounded system, the box size $L$ should be much larger than the maximal size of the flow fluctuations $L_\alpha$ (\ref{Lalpha}), $L\gg L_\alpha$. And the film thickness $h$ should be much less than the viscous scale $L_\nu$ (\ref{Lnu}), $h\ll L_\nu$.

\section{Coherent vortex}
\label{sec:coherent}

Now we turn to the case of large $L_\alpha$, $L_\alpha\gg L$. The condition can be achieved, say, by increasing the force power $\epsilon$, see Eq. (\ref{Lalpha}). Besides, we assume, that the box size $L$ is still much larger than the pumping length, $k_f L \gg 1$. Then there is the room for the inverse cascade carrying the energy from the pumping scale $k_f^{-1}$ to the scale $L$. However, the inverse cascade is essentially deformed in the case $L_\alpha\gg L$ in comparison with the unbounded system.

Since the energy, produced by the pumping force, cannot flow to scales larger than $L$, the energy flux is stopped there and the energy is accumulated until the incoming energy flux is compensated by the bottom friction. Then we arrive at the estimate $V\sim \sqrt{\epsilon/\alpha}$ for the velocity fluctuations at the scale $L$, following from the energy balance (\ref{energyb}). The quantity $\sqrt{\epsilon/\alpha}$ is much larger than the Kolmogorov estimate $(\epsilon L)^{1/3}$ (\ref{inversec}), since such velocity amplitude is characteristic of the scale $L_\alpha$ for the unbounded system, and $L_\alpha\gg L$.

The large-scale fluctuations can be chaotic, then we deal with the random large-scale motion \cite{DFKL22}. However, at some conditions coherent vortices appear that are long-living structures of the size of the order of the box size $L$. Such coherent vortices were observed both in laboratory experiments \cite{Shats09,Orlov17} and in numerical simulations \cite{Chertkov2007,Laurie2014,Parfenyev2022}. The coherent vortex possesses a well-defined mean velocity profile defined in the reference system attached to the vortex center. Note that the center of the vortex moves with some random velocity.

Based on the experimental and the numerical results, one can assert that the vortex is isotropic in average. By other words, one deals with a differential rotation, described by the polar velocity $U(r)$ where $r$ is the distance between the observation point and the center of the vortex. The mean vorticity of the coherent vortex $\varOmega =\partial_r U +U/r$ is a function of the distance $r$ as well. The quantity $\varOmega(r)$ is maximal at the center of the coherent vortex.

An equation for $U$ can be derived from the basic equation (\ref{basic0}) where one has to separate the mean flow and fluctuations on its background. Below, we designate as $\bm v$ the velocity of the fluctuations, unlike the previous sections. Taking the polar component of the equation (\ref{basic0}) and averaging it over time, one finds
 \begin{equation}
 \alpha U=-\left(\partial_r+\frac{2}{r}\right)\langle v_r v_\varphi \rangle
 +\nu \left(\partial_r^2 + \frac{1}{r}\partial_r -\frac{1}{r^2}\right) U,
 \label{Ueq}
 \end{equation}
where $v_r$ and $v_\varphi$ are radial and polar components of the velocity fluctuations. At deriving Eq. (\ref{Ueq}) we assumed that the mean pumping force is zero. The average $\langle v_r v_\varphi \rangle$ in Eq. (\ref{Ueq}) is no other than the off-diagonal component of the Reynolds stress tensor \cite{MoYa}.

The viscous term in the equation (\ref{Ueq}) is relevant in the region near the vortex center, the size of the region can be estimated as $\sqrt{\nu/\alpha}$. In this region, inside the vortex core, the viscous term dominates, forcing the solid state rotation $U\propto r$: such dependence turns to zero the viscous term in the equation (\ref{Ueq}). Outside the core the viscous term in the equation (\ref{Ueq}) is irrelevant. Then the mean profile $U$ is determined by the balance between the bottom friction and the force, related to the Reynolds stress tensor, generated by the fluctuations.

Separating in Eq. (\ref{basic0}) the mean flow and the fluctuations, one finds the following equation for the fluctuating velocity
\begin{eqnarray}
\partial_t v_\lambda
+{\lfloor(\bm v \nabla) v_\lambda \rfloor }
+\partial_\lambda p
\nonumber \\
+\frac{U}{r}\partial_\varphi v_\lambda
-\frac{U}{r}\varepsilon_{\lambda\mu}v_\mu
-v_r \varSigma \frac{\varepsilon_{\lambda\mu} r_\mu}{r}
\nonumber \\
=-\alpha v_\lambda  +\nu \nabla^2 v_\lambda +f_\lambda,
\label{basic3}
\end{eqnarray}
where ${\lfloor A \rfloor =A- \langle A \rangle}$, $\varepsilon_{\lambda\mu}$ is $2d$ antisymmetric Levi-Civita symbol and
\begin{equation}
\varSigma=r\partial_r(U/r)
=\partial_r U -U/r.
\label{avsigma}
\end{equation}
The quantity (\ref{avsigma}) is the local shear rate of the mean flow. Now pressure $p$ satisfies the equation
\begin{eqnarray}
\nabla^2 p=
-\partial_\lambda \partial_\mu \lfloor v_\mu  v_\lambda\rfloor 
\nonumber \\
+\frac{2}{r}\left[U \varpi + \varSigma(v_\varphi-\partial_\varphi v_r)\right],
\label{pressure2}
\end{eqnarray}
instead of Eq. (\ref{pressure}). {Here $\varpi=\mathrm{curl}\, \bm v$ -- fluctuating vorticity.}

Let us derive an analog of the energy balance (\ref{energyb}) for the interior of the vortex. Multiplying the equation (\ref{basic3}) by $\bm v$ and averaging, one finds
\begin{eqnarray}
\frac{1}{r}\partial_r\left[r\langle v_r
(p+v^2/2)\rangle \right]
+\varSigma \langle v_r v_\varphi \rangle
\nonumber \\
=\epsilon -\alpha \langle v^2\rangle
+\nu \langle \bm v \nabla^2 \bm v \rangle.
\label{energybal}
\end{eqnarray}
Note that the relation is correct even for an inhomogeneous forcing. Due to the random motions of the vortex the inhomogeneity is effectively averaged. Therefore in this case $\epsilon$ in Eq. (\ref{energybal}) is an average over the box.

{The equation for the fluctuating vorticity $\varpi$ inside the coherent vortex is
\begin{eqnarray}
\partial_t \varpi +\frac{U}{r}\partial_\varphi \varpi
+v_r \partial_r \Omega
+\lfloor (\bm v \nabla)\varpi \rfloor
\nonumber \\
=-\alpha \varpi  +\nu \nabla^2 \varpi +\phi,
\label{vorticfluc}
\end{eqnarray}
where $\varOmega=\partial_r U+U/r$ is the mean angular velocity and $\phi={\mathrm curl}\, \bm f$, as earlier. The equation (\ref{vorticfluc}) can be obtained by taking $\mathrm curl$ of the equation (\ref{basic3}) or directly from Eq. (\ref{basic}) by separating the mean and the fluctuating contributions to the flow velocity.
}

{Let us consider the separations $r$ from the vortex center much larger than the characteristic length of the fluctuations. Then the term with $\varOmega$ in Eq. (\ref{vorticfluc}) is negligible. Passing then to the reference system rotating with the angular velocity $U(R)/R$ we find from Eq. (\ref{vorticfluc}) for $r$, close to $R$
\begin{eqnarray}
\partial_t \varpi +\frac{\varSigma(R)}{R}(r-R)\partial_\varphi \varpi
+\lfloor (\bm v \nabla)\varpi \rfloor
\nonumber \\
=-\alpha \varpi  +\nu \nabla^2 \varpi +\phi.
\label{vorticfluc2}
\end{eqnarray}
We conclude that the influence of the mean flow on the flow fluctuations is reduced to the presence of the effective shear flow.
}

\section{Fluctuations inside a coherent vortex: quasilinear regime}
\label{sec:quasilin}

The mean flow suppresses the flow fluctuations inside the coherent vortex. That leads to diminishing the non-linear interaction of the fluctuations. We consider the distances $r$ satisfying $k_f r\gg1$. Then the fluctuations can be thought as living in the effectively shear flow with the shear rate (\ref{avsigma}). As it was demonstrated in Refs. \cite{KLT23a,KLT23b,KL24} the suppression effect is relevant, provided
\begin{equation}
\varSigma\gg \nu k_f^2, \quad
\varSigma^2 \gg \alpha^3/(\nu k_f^2).
\label{condition}
\end{equation}
The conditions (\ref{condition}) mean that the fluctuations of the size $k_f^{-1}$ produced by the pumping force, are essentially deformed by the shear flow before they are damped by dissipation (viscosity and bottom friction).

The analysis of the interaction corrections to the correlations functions of the fluctuating vorticity \cite{KLT23a,KL24} shows that the parameter governing the strength of the interaction of the fluctuations is
\begin{eqnarray}
\beta =\frac{\epsilon}{\nu \varSigma^2},
\label{parbeta}
\end{eqnarray}
at the conditions (\ref{condition}). The interaction of the fluctuations is weak provided $\beta$ is small. Note that the bottom friction coefficient $\alpha$ does not enter the expression for the perturbation parameter $\beta$ (\ref{parbeta}) though it figures in Eq. (\ref{condition}). {The parameter (\ref{parbeta}) can be understood as the ratio of $\epsilon/(\nu k_f^2)$ and $\varSigma^2 k_f^2$. The first factor is the square of the typical velocity produced by the pumping force in the presence of viscosity, and the second factor is the square of the velocity deformation, caused by the shear flow with the rate $\varSigma$ at the pumping scale. }

If the interaction is small then it is possible to neglect the nonlinear term {in the equation (\ref{vorticfluc2}) to obtain the linear equation for the fluctuating vorticity $\varpi$.} In the literature such situation is called quasilinear one. In the quasilinear approximation, one can consistently calculate correlation { functions of $\varpi$ solving the equation for $\bm v$ in terms of $\phi$ and averaging the corresponding products of $\varpi$ based on statistical properties of  $\phi$  \cite{KL15,kolokolov2016structure,kolokolov2016velocity,KL17,Frishman18}. One can find then correlation functions of the velocity. Corrections to the correlation functions related to the non-linear interaction of the flow fluctuations were examined in Refs. \cite{KLT23b,KL24}. }

Since the nonlinear interaction of the fluctuations is irrelevant in the quasilinear regime, then the direct and the inverse cascades are absent in the case inside the coherent vortex. The size of the flow fluctuations, produced by the pumping force, can be estimated as $k_f^{-1}$ in the radial direction. However, their characteristic size in the angular direction is much larger since the mean flow elongate the fluctuations in the angular direction. The degree of the elongation depends on the life time of the fluctuations. Say, the pair correlation function of vorticity is formed during the time
\begin{equation}
\tau_\star =\left(\varSigma^2 \nu k_f^2\right)^{-1/3}.
\nonumber
\end{equation}
Therefore the characteristic angular size of the fluctuations is
\begin{equation}
\varSigma \tau_\star k_f^{-1}
=\left(\frac{\varSigma}{\nu k_f^2}\right)^{1/3}k_f^{-1}.
\nonumber
\end{equation}
The size is much larger than the radial size thanks to the first condition in Eq. (\ref{condition}).

One of the results of the calculations in the quasilinear approximation concerns the correlation function $\langle v_r v_\varphi \rangle$ figuring in Eq. (\ref{Ueq}). As it was demonstrated in Refs.   \cite{KL15,kolokolov2016structure}
\begin{equation}
\langle v_r v_\varphi \rangle= \epsilon/\varSigma.
\label{vrvvarphi}
\end{equation}
Substituting the expression (\ref{vrvvarphi}) into Eq. (\ref{Ueq}) we find a closed equation for the mean velocity $U$, since $\varSigma$ is expressed via $U$ in accordance with Eq. (\ref{avsigma}). Outside the viscous core, where the term with the viscosity in Eq. (\ref{Ueq}) can be neglected, one finds the solution of the equation
\begin{equation}
U=\sqrt{3\epsilon/\alpha},
\label{flatpr}
\end{equation}
independent of $r$. The flat velocity profile (\ref{flatpr}) was observed in numerical simulations \cite{Laurie2014}, where the quasilinear regime was held. Note that the expression (\ref{flatpr}) is in accordance with the estimate $(\epsilon/\alpha)^{1/2}$ for the large-scale fluctuations found in Section \ref{sec:coherent}.

Calculating the mean vorticity $\varOmega$ corresponding to the flat velocity profile (\ref{flatpr}) we find $\varOmega= \sqrt{3\epsilon/\alpha}\, r^{-1}$. Thus the vorticity grows as $r$ diminishes and at small $r$ becomes much larger than the typical vorticity of the large-scale fluctuations. One can say that the coherent vortices accumulate vorticity. The law $\varOmega\propto r^{-1}$ works down to the viscous core where $\varOmega$ is saturated.

One can explain the relation (\ref{vrvvarphi}) in simple terms based on the energy balance (\ref{energybal}). The main contribution to the average $\langle v_r v_\varphi \rangle$ is related to the fluctuations with the radial scales of the order of the pumping length. Thus $\nabla^2$ in Eq. (\ref{pressure2}) is estimated as $k_f^2$. Therefore pressure $p$ is small at the condition $k_f r \gg1$ and can be neglected in Eq. (\ref{energybal}). The term, proportional to $v_r v^2$ in Eq. (\ref{energybal}) can be neglected as well, since it is of third order in $\bm v$ and is small in the quasilinear regime. Outside the viscous core one can neglect the viscous term in Eq. (\ref{energybal}). Since the fluctuations are weak, one can neglect the term with $\alpha$ in Eq. (\ref{energybal}) as well. Thus, one arrives at the relation (\ref{vrvvarphi}).

In accordance with Eq. (\ref{avsigma}), the shear rate corresponding to the velocity (\ref{flatpr}) is
\begin{equation}
\varSigma=-\frac{\sqrt{3\epsilon/\alpha}}{r}.
\nonumber
\end{equation}
Substituting the expression into Eq. (\ref{parbeta}) one finds
\begin{equation}
\beta =\frac{\alpha r^2}{3\nu}.
\label{betardep}
\end{equation}
The condition of smallness of such $\beta$ is compatible with the inequality $k_f r \gg1$ only if $\alpha \ll \nu k_f^2$. The inequality cannot be achieved in thin fluid films (see Section \ref{sec:general}), however, it is achievable for suspended soap or smectic films. There is no problem to satisfy the inequality $\alpha \ll \nu k_f^2$ in numerical simulations. The inequality was satisfied in the simulations reported in Ref. \cite{Laurie2014}.

\section{Fluctuations inside a coherent vortex: strong interaction}
\label{sec:strong}

Here we consider the case where the coupling constant $\beta$ (\ref{parbeta}) is large. Then the nonlinear interaction of the flow fluctuations inside the coherent vortex becomes relevant. Note that in accordance with the expression (\ref{betardep}) the coupling constant $\beta$ grows as $r$ increases. Therefore one can encounter the situation where both regimes (the quasilinear one and the regime of strong interaction) coexist inside a single coherent vortex.

Since the interaction of the fluctuations becomes relevant for large $\beta$, the direct and the inverse cascades are restored unlike the quasilinear regime. However, the correlation functions of the velocity and vorticity characteristic of the cascades can be anisotropic in some regions of scales. It is a direct consequence of the elongation of the flow fluctuations in the angular direction produced by the mean flow $U$ of the vortex.

Let us consider the inverse cascade inside the vortex {One can extend Kolmogorov arguments \cite{Kol41,LL87} for the case. As it follows from Eq. (\ref{vorticfluc2}), the effective shear flow is dropped from the generalized relation (\ref{inverseflux}) if $r_1=r_2=r$.} Therefore we arrive at the estimate
\begin{equation}
{v_\varphi(r,\varphi_1)-v_\varphi(r,\varphi_2)}
\sim (\epsilon r |\varphi_1-\varphi_2|)^{1/3},
\label{kolomogangle}
\end{equation}
{formally coinciding with the isotropic estimate (\ref{inversec}).}

{Comparing the shear rate $\varSigma$ with the nonlinear rate $r^{-1}\partial _\varphi v_\varphi$, we find using the expression (\ref{kolomogangle}) the scale}
\begin{equation}
L_{an}=\epsilon^{1/2}\varSigma ^{-3/2},
\label{kolmo5}
\end{equation}
separating the isotropic regime and the anisotropic one. For scales smaller than $L_{an}$ the mean shear flow is irrelevant and the standard isotropic inverse cascade is realized whereas for scales larger than $L_{an}$ the mean shear flow becomes relevant. Thus, for scales larger than $L_{an}$ the inverse cascade becomes anisotropic. If $k_f L_{an}<1$ then the inverse cascade is anisotropic on all scales above the pumping length.

Let us establish the character of the anisotropy. {For the purpose one has to compare the term with $\varSigma$ and the nonlinear term in the equation (\ref{vorticfluc2}). As a result, we find for the second-order structure function
\begin{equation}
\langle [v_\varphi(r_1,\varphi_1)
-v_\varphi(r_2,\varphi_2)]^2\rangle
=[\epsilon r (\varphi_1-\varphi_2)]^{2/3} g(\xi),
\label{pairpolar}
\end{equation}
where $r=r_1/2+r_2/2$, $g$ is some dimensionless function and $\xi$ is the similarity variable
\begin{equation}
\xi=\varSigma (r_1-r_2)
[\epsilon r (\varphi_1-\varphi_2)]^{-1/3}.
\label{selfsimil}
\end{equation}
Analogously, higher order structure functions of $v_\varphi$ can be expressed.}

The incompessibility condition $\nabla \bm v=0$ leads to the conclusion {that $\Delta r \Delta v_\varphi \sim r \Delta \varphi \Delta v_r$ where we introduced the differences like in Eqs. (\ref{pairpolar},\ref{selfsimil}). If the similarity variable $\xi\sim 1$, then we conclude that $\Delta v_r \ll \Delta v_\varphi$ due to $\Delta r \gg L_{an}$, as one could anticipate. Based on the reasoning one can establish the scaling laws like (\ref{pairpolar}) for the structure functions of any order containing both, $v_\varphi,v_r$.   }

Note that if the similarity variable (\ref{selfsimil}) is of order of unity, $\xi\sim1$, then
\begin{equation}
\frac{\Delta r}{r \Delta\varphi}\sim
\left(\frac{L_{an}}{r\Delta \varphi}\right)^{2/3}.
\label{anisotropy}
\end{equation}
The quantity (\ref{anisotropy}) is small for the region of the anisotropic inverse cascade, where $r\Delta \varphi\gg L_{an}$. By other words, the characteristic separation in the angular direction is much larger than in the radial one, as one expected.

{The anisotropic scaling (\ref{pairpolar}) is correct provided $|\varphi_1-\varphi_2|\ll1$. If the difference is of order unity then $\xi\sim1$ leads to
\begin{equation}
\Delta r \sim r^{1/3} L_{an}^{2/3} \ll r .
\label{finishic}
\end{equation}
If $\Delta r $ becomes larger than the estimate (\ref{finishic}) then the term with $\varSigma$ in the equation (\ref{vorticfluc2}) overcomes the nonlinear term and we pass to the quasilinear regime, where the energy cascade is absent. Our notion of the regime \cite{KLT23b,KL24} enables us to conclude that the quantity (\ref{finishic}) is the correlation length of the flow fluctuations in the radial direction. Therefore the strongest velocity fluctuations can be estimated as
\begin{equation}
v_\varphi \sim (\epsilon r)^{1/3}, \quad
v_r \sim \frac{L_{an}^{2/3}}{r^{2/3}} (\epsilon r)^{1/3},
\label{finishic2}
\end{equation}
in accordance with Eq. (\ref{finishic}).
}

Now we return to the energy balance (\ref{energybal}). {As it follows from Eq. (\ref{pressure2}), the first contribution to pressure $p$ can be estimated as $\bm v^2$ where $\bm v$ is the strongest velocity fluctuation determined by Eq. (\ref{finishic}). The second contribution to pressure $p$ related to the last two terms in the right hand side of Eq. (\ref{pressure2}) is negligible since Laplacian in this case is estimated as $(\Delta r)^{-2}$, where $\Delta r$ is determined by Eq. (\ref{finishic}). Therefore the Laplacian produces a big factor at $p$. Thus
\begin{equation}
v_r(p+v^2/2) \sim
\frac{L_{an}^{2/3}}{r^{2/3}}\epsilon r,
\nonumber
\end{equation}
in Eq. (\ref{energybal}). The derivative $\partial_r$ in the term with pressure in Eq. (\ref{energybal}) is estimated as $r^{-1}$ and therefore this term contains the small factor $(L_{an}/r)^{2/3}$ at $\epsilon$ and can be, consequently, neglected.
}

Thus we arrive at the same relation (\ref{vrvvarphi}) and to the same flat velocity profile (\ref{flatpr}) as in the quasilinear approximation. The profile is realized provided there is an anisotropic region of the inverse cascade. Note that for the flat velocity profile
\begin{equation}
L_{an}^2 \sim r^3/L_\alpha^2 \ll r.
\nonumber
\end{equation}
Therefore the expression (\ref{flatpr}) for the mean velocity is self-consistent.

Now some words about the direct (enstrophy) cascade. If $k_f L_{an}\gg 1$ then the direct cascade is the same as in the unbounded system, see Section \ref{sec:unbounded}. In the opposite case, where $k_f L_{an}\ll 1$ the direct cascade is anisotropic. The inequality is equivalent to $\varSigma^3\gg \eta$. {Analogously to the inverse cascade, $r\Delta \varphi \gg \Delta r$ for the characteristic values. Thus $\varpi= -\partial_r v_\varphi$. Comparing then the term with $\varSigma$ and the nonlinear term in the equation (\ref{vorticfluc2}) we find the estimates
\begin{eqnarray}
\varpi \sim \varSigma, \quad v_\varphi \sim \Sigma \Delta r.
\label{directan1}
\end{eqnarray}
As in the unbounded case, the characteristic $\varpi$ is independent of scale.

Expanding the Kolmogorov-Kraichnan reasoning \cite{67Kra,68Lei,69Bat} to the anisotropic direct cascade we find the relation
\begin{equation}
\langle \Delta v_\varphi \varpi \varpi \rangle
\sim \eta r \Delta \varphi,
\label{directan2}
\end{equation}
for $\Delta r=0$. Substituting the estimates (\ref{directan1}) into Eq. (\ref{directan2}) we find the similarity variable
\begin{equation}
(\Sigma^3\Delta r)/(\eta r\Delta \varphi),
\label{directan3}
\end{equation}
controlling correlation functions of vorticity in the direct cascade.  }

\section{Conclusion}
\label{sec:conclu}

Two-dimensional turbulence is in some sense reacher than three-dimensional one. There are two cascades (of energy and of enstrophy) leading to production of a large diapason of scales of turbulent pulsations. As a consequence of the energy cascade, there can appear coherent vortices in a finite box. We examined different regimes of the flow fluctuations in a coherent vortex, that can be quasilinear or non-linear. In the last case the anisotropic scaling has to be observed.

The theoretical results presented in the paper are mainly confirmed by laboratory experiments with thin fluid films and by numerical simulations. However, the theoretical results concerning the anisotropic scaling in the non-linear regime wait for their confirmation. The numerical simulations initiated to check our predictions are now in progress. We are thinking about experimental check as well.

We considered the simplest model where both, the box and the pumping force are homogeneous. We believe that the model demonstrates all qualitative features of two-dimensional turbulence. However one can expand the model to include into consideration inhomogeneity of the box and of the pumping force. The results of such model can be in more detail compared to hydrodynamic processes in environment.

One should note such phenomenon as geostrophic vortices, generated in a relatively fast rotating fluid, and playing an essential role in geophysics \cite{Pedlovsky}. The geostrophic vortices can be described in terms of an effectively two-dimensional flow governed by the two-dimensional hydrodynamic equations \cite{Ogorodnikov20}. It would be of interest to translate our results concerning the anisotropic scaling to the geostrophic vortices.

Another possible direction of expansion of our theoretical scheme is passing to non-Newtonian fluids. Particularly, one can think about polymer solutions. Effects caused by the elastic degree of freedom related to the polymers could lead to such remarkable phenomenon as elastic turbulence \cite{Steinberg}. It would be of interest to examine peculiarities of elastic turbulence in thin fluid films.

\acknowledgments

We thank V. Parfenyev for numerous helpful discussions. The work is performed in the Laboratory ``Modern Hydrodynamics'' created in frames of Grant 075-15-2022-1099 of the Ministry of Science and Higher Education of the Russian Federation in Landau Institute for Theoretical Physics of RAS and is supported by Grant 23-72-30006 of Russian Science Foundation.

\end{document}